\documentclass[aoas,MSNbibl,nameyear,seceqn,dvips]{arximspdf}
\usepackage{dcolumn}
\usepackage{graphicx}

\doi{10.1214/13-AOAS679} 
\volume{7}
\issue{4}
\pubyear{2013}
\firstpage{2106}
\lastpage{2137}

\makeatletter
\newcolumntype{d}[1]{D{.}{.}{#1}}
\renewcommand{\mid}{\vert}
\makeatother

\begin{document}
\begin{frontmatter}

\title{Flexible covariate-adjusted exact tests of~randomized treatment
effects with~application~to~a~trial of HIV education\thanksref{TT1}}
\runtitle{Flexible covariate-adjusted exact inference}
\thankstext{TT1}{Supported by NIH Grants R37 51164 and T32AI007358.}

\begin{aug}
\author[A]{\fnms{Alisa J.} \snm{Stephens}\corref{}\ead[label=e1]{ajsteph@post.harvard.edu}\thanksref{t1}},
\author[B]{\fnms{Eric J.} \snm{Tchetgen Tchetgen}\thanksref{t2}}
\and\break
\author[B]{\fnms{Victor} \snm{De Gruttola}\thanksref{t2}}
\runauthor{A. J. Stephens, E. J. Tchetgen Tchetgen and V. De Gruttola}
\affiliation{University of Pennsylvania\thanksmark{t1} and Harvard
University\thanksmark{t2}}
\address[A]{A. J. Stephens\\
Center for Clinical Epidemiology and Biostatistics\\
Perelman School of Medicine\\
University of Pennsylvania\\
423 Guardian Drive\\
624 Blockley Hall\\
Philadelphia, Pennsylvania 19103\\
USA\\
\printead{e1}}
\end{aug}
%
\address[B]{E. J. Tchetgen Tchetgen\\
V. De Gruttola\\
Harvard University\\
Boston, Massachusetts 02115\\
USA}

\received{\smonth{6} \syear{2012}}
\revised{\smonth{8} \syear{2013}}

%
\begin{abstract}
The primary goal of randomized trials is to compare the effects of
different interventions on some outcome of interest. In addition to the
treatment assignment and outcome, data on baseline covariates, such as
demographic characteristics or biomarker measurements, are typically
collected. Incorporating such auxiliary covariates in the analysis of
randomized trials can increase power, but questions remain about how to
preserve type~I error when incorporating such covariates in a flexible
way, particularly when the number of randomized units is small. Using
the Young Citizens study, a cluster-randomized trial of an educational
intervention to promote HIV awareness, we compare several methods to
evaluate intervention effects when baseline covariates are incorporated
adaptively. To ascertain the validity of the methods shown in small
samples, extensive simulation studies were conducted. We demonstrate
that randomization inference preserves type~I error under model
selection while tests based on asymptotic theory may yield invalid
results. We also demonstrate that covariate adjustment generally
increases power, except at extremely small sample sizes using liberal
selection procedures. Although shown within the context of HIV
prevention research, our conclusions have important implications for
maximizing efficiency and robustness in randomized trials with small
samples across disciplines.
\end{abstract}

%

\end{frontmatter}

\section{Introduction}

The Young Citizens study was a cluster-randomized trial designed to
evaluate the impact of involving adolescents in a role-play
intervention on HIV awareness and education. In addition to the primary
outcome of child efficacy, a score reflecting the degree to which
community members believe adolescents can effectively educate their
families and peers, the study collected extensive demographic data that
described the clusters (actual communities in \mbox{Tanzania}) and the
individuals within each community who participated in the study. To
protect type~I error when evaluating the effect of the randomized
intervention, current practice requires prespecification of the methods
for including baseline community-level and individual-level covariates
in analyses, whether as stratification factors or as control covariates
in a regression model. Recently developed methods allow for more
flexible model selection characterizing the outcome-baseline covariate
relationship without loss of protection of type~I error, at least
asymptotically. Several studies have demonstrated the value of new
methods in permitting flexible use of baseline correlates of the
outcome to improve power and efficiency in treatment effect estimation
[\citet{Tsiatis08}, \citet{Tsiatis08b}, \citet{Stephens}]. These methods, however,
rely on asymptotic arguments that may not apply to studies like Young
Citizens that have a small number of randomized units. For such
studies, additional variability introduced by flexible model selection
may result not only in failure to preserve type~I error but also loss
of power and efficiency as compared to unadjusted analyses. Motivated
by Young Citizens, which had only 15 clusters per arm, we evaluate
several flexible covariate adjustment methods for studies with small
numbers of randomized units and large numbers of potential adjustment
variables. We apply each method to Young Citizens data and report on
simulation studies conducted to compare power and the degree of
protection of type~I error among tests.

In a randomized trial like Young Citizens, researchers typically
measure data on outcomes, baseline covariates and the treatment
assignment. Although abundant baseline covariate data are often
available, the primary analysis is often a comparison of outcomes among
subjects assigned to different levels of treatment without
consideration of covariates. For scalar outcomes, tests comparing some
\mbox{feature} of the outcome distribution under treatment versus control are
used to assess the statistical significance of observed differences in
outcomes across treatment groups. When outcomes are multivariate, as in
Young Citizens, modified versions of these tests are available to
adjust standard errors for correlation among multiple measurements
within the same randomized unit [\citet{Donner}]. Any collection
of baseline covariates potentially explains variability in outcomes,
and \mbox{incorporating} them in analyses may therefore increase efficiency.

A variety of methods are available to incorporate baseline covariates
in trial analyses. Regression analysis is one approach that may be used
to estimate and test for treatment effects and, in some cases, to
permit covariate adjustment that guarantees efficiency improvement over
unadjusted analyses. We first discuss models that ignore baseline
covariates, and then compare them to models that adjust for\vadjust{\goodbreak} baseline
covariates. Ignoring baseline covariates, the effect of a binary
treatment on the marginal mean outcome may be assessed from a
generalized linear model with treatment as a predictor; such a model is
commonly referred to as the marginal treatment model. Model parameters
can be estimated using semiparametric estimating equations or fully
parametric maximum likelihood inference. The weak null hypothesis of
equal mean outcomes for the intervention groups, also known as Neyman's
null hypothesis, is tested using the estimated treatment coefficients.
Under randomization, this test is equivalent to a test of no average
causal effect of treatment. Estimation strategies are available to
accommodate multivariate, \mbox{dependent} outcomes. Baseline covariates are
often incorporated by assuming a conditional mean model (CMM) to obtain
inferences on the conditional effect of treatment. For models with an
identity link function, such as linear models, when the true model does
not contain any treatment--covariate interactions, independence of
intervention and covariates (which follows from randomization)
guarantees that the adjusted estimator is consistent for the marginal
treatment effect and has lower variance than does the unadjusted
estimator, even under misspecification of the model's covariate
functions [\citet{Tsiatis08}]. For other link functions, the
addition of baseline covariates to the assumed mean model does not
guarantee variance reduction.

\citeauthor{Tsiatis08b} (\citeyear{Tsiatis08b}) introduced covariate
adjustment with asymptotically guaranteed efficiency improvement for
general link functions in a class of augmented estimators. Augmented
estimators are derived from semiparametric theory through augmenting
standard estimating functions by the \mbox{subtraction} of their Hilbert space
projection onto the span of the scores of the treatment mechanism.
Semiparametric theory provides theoretical justification for the
efficiency improvement of augmented estimators in large samples under
the assumed marginal model, irrespective of the link function.
\citet{Stephens} demonstrated the use of
such estimators applied to clustered or longitudinal data by augmenting
Generalized Estimating Equations (GEE). The same authors also presented
the locally efficient augmented estimator under the marginal treatment
model [\citet{Stephens2}]. Augmented inference relies on
asymptotic theory; for large samples, model selection variability for
baseline covariates is small when the number of covariates is small as
well. When the number of randomized units is small, however, flexible
covariate selection induces additional variability that may lead to
efficiency loss and underestimation of standard errors. To evaluate the
intervention effect in Young Citizens, we therefore require analytical
strategies that are valid in small samples.

To avoid reliance on asymptotic theory, \citeauthor{Rosenbaum}
(\citeyear{Rosenbaum}) extended the randomization theory of
\citeauthor{Fisher35} (\citeyear{Fisher35}) to propose an exact
covariate-adjusted test that does not assume a particular distribution
for outcomes or that the observed data are a random sample from some
unobserved population of independent units. Randomization inference
considers a subject's potential outcomes under each treatment level.
Under the so-called consistency assumption in the potential outcomes
framework, a subject's observed outcome is equal to his or her
potential outcome under the treatment that he or she actually received.
Consistency requires (1) that there is a single version of the
intervention in view, so that it is well defined, and (2)~that there is
no interference between individuals, so that a person's exposure can
only affect his or her outcome. Potential outcomes under the treatment
not received are unobserved. Randomization tests condition on baseline
covariates and outcomes and test the strong null hypothesis of no
treatment effect on any individual's outcome, often referred to as
Fisher's null hypothesis, which amounts to equality of a subject's
potential outcomes under all possible treatments.
\citeauthor{Rosenbaum} (\citeyear{Rosenbaum}) discussed the potential
outcomes framework in detail. The null distribution of the test
statistic is obtained through permutation of the treatment among
randomized subjects. The test proposed by \citeauthor{GailTan}
(\citeyear{GailTan}) approximates the exact test by standardizing the
observed test statistic by its randomization-based variance. Post
model-selection inference based on the Gail et al. and Rosenbaum
approaches has not been investigated; below, we consider model
selection to determine covariates that explain variability in child
efficacy. Such adaptive selection of baseline covariates may be
particularly useful when the set of baseline covariates is
high-dimensional or prior knowledge is not available to inform
covariate adjustment. Further improvement in small-sample inference may
be possible from higher-order approximations of the distribution of
a~class of randomization test statistics [\citet{Bickel78}], but
this theory has not yet been evaluated in practice.

We consider four covariate-adjusted methods to test for an effect of
the role-play intervention on child efficacy in Young Citizens: (I)
conditional mean models (CMM), (II) marginal model with augmentation,
(III) approximate exact, and (IV)~exact (permutation), with details
discussed in Section~\ref{section:Methods}. Although the Young
Citizens outcomes are correlated within communities, we also present
inference for independent outcomes. These independent outcome methods
are relevant for studies involving rare diseases such as lymphomas or
leukemia, which typically have relatively few subjects, or in analysis
of clustered data based on average outcomes for each cluster. In
Section~\ref{section:Sim} the small sample properties of
covariate-adjusted tests are evaluated through simulation. Section~\ref{section:discussion} provides a summary of our results and
recommendations for practical use.

\section{The Young Citizens study}
Young Citizens was a cluster-randomized trial designed to evaluate the
effectiveness of an educational role-play intervention in training
adolescents to be peer educators about HIV transmission dynamics.
Thirty communities were randomized to intervention or control,\vadjust{\goodbreak}
resulting in 15 communities per arm. In communities randomized to
intervention, adolescents age 10--14 were selected to participate in
learning and performing a skit in which each participant assumed the
role of an agent involved in HIV transmission and genetic evolution.
Residents in intervention and control communities were surveyed and
asked to report the degree to which they believed adolescents could
effectively communicate to their families and peers about HIV. The
number of residents surveyed (cluster sizes) varied between 16 and 80
according to the size of the community's population. Data collected
included the child efficacy outcome---a child empowerment score
derived from individuals' responses to multiple survey items, the
cluster-level intervention assignment indicator, and various
demographic and household characteristics. Among the cluster-level
covariates were population density and designation of the community as
urban or rural. Covariates measured at the level of the individual
included household wealth, the number of adults in the house, the
number of children in the house, years spent in the current residence,
age and gender of the head of the household, and several wealth
indicators such as whether the house had a flushing toilet, electricity
or if the family in residence owned their own transportation. These
variables were summed to create a wealth score, which was then averaged
to calculate a community's mean wealth. Only one subject was surveyed
per household. Demographic characteristics such as religion and
employment status were also collected; indicators for home ownership,
knowledge of the local leader and number of relatives in the
neighborhood served as measures of the degree to which household
members were rooted in the community. The number of relatives in the
neighborhood further conveyed this information. A total of 1100
individuals were surveyed across all thirty communities, and data on
over 20 covariates were available for covariate adjustment.

\section{Methods}
\label{section:Methods} We consider four methods of covariate-adjusted
hypothesis testing to determine the impact of the HIV/AIDS education
intervention on child efficacy in \textit{Young Citizens}: (I) Wald
test of $\beta_1^*$ in the conditional treatment model, (II)~Wald test
of $\beta_1$ in the marginal treatment model, in which estimating
equations are augmented to include baseline covariates, (III)
approximate exact test, and (IV)~exact test. This list is not
comprehensive, but does include widely-recognized classical and modern
methods. We first present each test for independent outcomes and then
describe generalizations for dependent outcomes that allow correlation
among individuals within communities. Methods for independent outcomes
may be used with dependent outcome data under an analysis strategy that
averages individual level child efficacy scores and baseline covariates
within communities into a single community-level score for each
variable. In the third subsection, we present model selection methods
to identify the characteristics of communities and households that
correlate with child efficacy in order to enhance power in testing of
intervention effects.

In defining each method, we consider $n$ independent and identically
distributed units $O_i=(\mathbf{Y}_i,A_i,\mathbf{X}_i)$ chosen from a
population. For Young Citizens, the vector $\mathbf{Y}_i$ represents
the set of perceived child efficacy scores calculated from the surveys
of individuals within a community; more generally, $\mathbf{Y}_i$ is
the set of responses of trial participants within the same randomized
group, and $Y_{ij}$ is the response of the $j$th person from the $i$th
community. Similarly, in a longitudinal study, $\mathbf{Y}_i$ would
denote a set of repeated measurements on a single randomized subject,
whereas $Y_{ij}$ would reflect the $i$th person's outcome at the $j$th
time point. We consider settings where outcomes are vectors and the
treatment assignment is a scalar shared by responses within the same
cluster or subject. When presenting the simpler case of a single scalar
outcome for each randomized unit, $Y_i$ denotes the $i$th community's
average outcome. Young Citizens evaluates a binary role-play
intervention~$A_i$, but, more generally, $A_i=1,\ldots,K$ may represent
allocation to $1$ of $K$ possible treatment. Finally, $\mathbf{X}_i$ is
the set of baseline covariates, containing community-level
characteristics and individual-level measures. Individual-level
baseline covariates ${X}_{ij}$ are also averaged within community into
a single community score $X_i$ in analyses using methods for
independent data.

\subsection{Independent outcomes}\label{subsection:Ind}

\subsubsection{\texorpdfstring{Method \textup{Ia}: Wald test of $\beta_1^*$ in the conditional treatment model}
{Method \textup{Ia}: Wald test of beta1* in the conditional treatment model}}
Perhaps the most widely used method of covariate adjustment assumes a
conditional mean model specifying how mean values of $Y_{i}$ vary with
baseline covariates $\mathbf{X}_{i}$ and intervention $A_i$ up to an
unknown parameter $\beta$. Applying this method to cluster-averaged
Young Citizens data, we test the effect of the role-play on average
community mean child efficacy, conditional\vspace*{-2pt} on covariates, by evaluating
$H_0\dvtx \beta_1^* =0$ and\vspace*{-2pt} calculating the test statistic $T_c =
\frac{\hat{\beta}{}_1^*}{\widehat{\operatorname{SE}}(\hat{\beta}{}_1^*)}$. This
approach is standard in all statistical software packages.

\subsubsection{\texorpdfstring{Method \textup{IIa}: Wald test of $\beta_1$ in
the marginal model with augmented estimating equations [\citet{Tsiatis08}, \citet{Tsiatis08b},
\citet{vdljamie}]}
{Method \textup{IIa}: Wald test of beta1 in
the marginal model with augmented estimating equations [Tsiatis et~al. (2008),
Zhang, Tsiatis and Davidian (2008), van~der Laan and Robins (2003)]}}

Unlike inference based on the CMM, the augmentation method assumes the
less restrictive marginal model. Household and community covariate
information are captured by incorporating predicted values from a
conditional working mean model $E[Y_i\mid
\mathbf{X}_i,A_i=a]=d(\mathbf{X}_i;\eta_a)$ in estimating equations for
$\beta$. Consistent estimates of the marginal intervention effect
$\beta_1$ are obtained even if the working mean model is misspecified,
following from the double robustness property and the fact that the
treatment distribution is known [\citet{vdljamie}].

The null hypothesis of no effect of intervention on the average
community mean response ($H_0\dvtx \beta_1=0$) marginalizing over
covariates is tested by the statistic
\mbox{$T_a=\frac{\hat{\beta}_1}{\widehat{\operatorname{SE}}(\hat{\beta}_1)}$}, where
$\hat{\beta}_1$ is the solution of the augmented estimating equations
\begin{eqnarray*}
&& \sum_{i=1}^n\psi_{a}(O_i;
\beta)
\\
&&\qquad = \sum_{i=1}^n
\Biggl[h(A_i;\beta) \bigl\{ Y_i-g(A_i;
\beta) \bigr\}
\\
&&\hspace*{50pt} {} - \sum_{a=1}^K \bigl
\{I(A_i=a)-\pi_a \bigr\} \bigl\{h(a;\beta)
\bigl(E[Y_i\mid \mathbf{X}_i,A_i=a]
-g(a;\beta) \bigr) \bigr\} \Biggr]
=\mathbf{0}
\end{eqnarray*}
for any conformable function of treatment $h(A_i;\beta)$ and $\pi_a$
the probability of assignment to treatment $a$. As implied by the
subscript $a$, the regression for \mbox{augmented} estimators conditions on
the intervention assignment. To enhance objectivity, working
conditional models may be estimated separately in each treatment arm,
resulting in $K$ regression models that do not contain indicators for
\mbox{treatment.} The variance of $\hat{\beta}_1$ is estimated by the sandwich
variance estimator
\[
\widehat{\operatorname{Var}}(\hat{\beta}_1)=C \Biggl[ \Biggl(\sum
_{i=1}^n\frac{d h(A_i;\beta)}{d\beta^{\mathrm{T}}}{
D_i} \Biggr)^{-1^{\mathrm{T}}}\sum_{i=1}^n{
\bigl[\psi_{a}(O_i;\beta)^{\otimes2} \bigr]} \Biggl(
\sum_{i=1}^n\frac{d h(A_i;\beta)}{d\beta^{\mathrm{T}}}{
D_i} \Biggr)^{-1}\Biggr],
\]
where
\[
{D}_i=\frac{dg(A_i;\beta)}{d\beta^{\mathrm{T}}},\qquad U^{\otimes2}=UU^{\mathrm{T}}
\]
and
\[
C =\bigl\{(n_0-p_0-1)^{-1}+(n_1-
p_1-1)^{-1}\bigr\}/ \bigl\{(n_0-1)^{-1}+(n_1-1)^{-1}
\bigr\}
\]
is incorporated to account for finite-sample variability attributable
to augmenting [\citet{Tsiatis08}]. In C, $n_a$ is the sample size in
treatment arm $a$, and $p_a$ is the dimension of $\eta_a$ for the
working covariate-adjustment model.

\subsubsection{Method \textup{IIIa}: Approximation of the exact test
[\texorpdfstring{\citet{GailTan}}{Gail, Tan and Piantadosi (1988)}]}

The approximate exact test considers the null hypothesis $H_0\dvtx
y_{a}=y^{*}$ for all $a$, interpreted as no effect of intervention on
any Young Citizens community's mean response. This hypothesis is
stronger than the mean null assumption of no effect of intervention on
average community mean responses tested in Ia~and~IIa. To test $H_0$,
we construct the statistic
\[
T_s = \frac{S}{\sqrt{\operatorname{Var}(S\mid Y,\mathbf{X})}},\qquad\mbox{where } S=\sum
_{i=1}^n{(A_i-\pi)w_i},
\]
$\pi$ is the probability of assignment to the intervention $(A_i=1)$
arm, and \mbox{$\operatorname{Var}(S\mid Y,\mathbf{X})$} is shown in
(\ref{eq:variance}). Baseline covariates are incorporated by setting
$W_i=\hat{\varepsilon}_i=Y_i-d(\mathbf{X}_i;\hat{\eta})$, the residual
from the working mean model $E[Y_i\mid \mathbf{X}_i]=
d(\mathbf{X}_i;\eta)$ for a known function $d(\cdot)$ and estimated
parameter $\eta$. We omit the subscript $a$ on the regression function
as a reminder that under the strong null, $Y_i$ cannot depend on
treatment. The intervention is therefore excluded from the working
model. For unadjusted analysis, $W_i=Y_i$. In the following definition,
we use lowercase $w_i$ to reflect conditional inference based on $y_i$
and $\mathbf{x}_i$, the observed values of ${Y_i}$ and $\mathbf{X}_i$,
respectively. The variance of $S$ is calculated by
%
%
\begin{equation}
\label{eq:variance} \operatorname{Var}(S\mid Y,\mathbf{X})=\pi(1-\pi)\sum
_{i=1}^n{w_i^2}+
\overbrace{ \biggl(\pi\frac{n/2-1}{n-1}-\pi^2 \biggr)\sum
_{i\neq i'}w_iw_{i'}}^{Q}
\end{equation}
and significance is determined by comparing $|T_a|$ to the standard
normal distribution.

Term $Q$ in $\operatorname{Var}(S\mid \mathbf{Y},\mathbf{X})$ is nonzero
when the total number of subjects assigned to each intervention is
fixed. A fixed randomization scheme was used in Young Citizens and is
customary in trials with small samples, where matching and blocked
randomization strategies are employed to prevent imbalances in
treatment \mbox{allocation.} The vector $\mathbf{A}=(A_1,A_2,\ldots,A_n)$ then
follows a hypergeometric distribution, where the probability of being
assigned to treatment for a particular randomized unit is affected by
other units' treatment assignments. Independence of $\varepsilon_i$ and
$E[\varepsilon_i\mid \mathbf{X}_i]=0$ result in $Q \approx0$ when $W_i$
is a residual. If considering the unadjusted outcomes $Y_i$ in small
samples, failure to include $Q$ may result in gross variance
overestimation and conservative testing. In large samples, $Q \approx
0$ for either definition of $W_i$.

For the class of statistics defined by $T=\sum_{i=1}^n{A_ic_i}$, where $c_i$
is a score, \citeauthor{Bickel78} (\citeyear{Bickel78}) determined a
higher-order approximation for the permutation conditional distribution
of the standardized statistic $T^{*}$, given by
\[
P \bigl(T^{*}<t \bigr)=\Phi(t)-\frac{\phi(t)}{\pi(1-\pi)}
\bigl[C_1H_1(t)+ C_2H_2(t)+
C_3H_3(t)+C_5H_5(t) \bigr],
\]
where $H_1(t)-H_5(t)$ and $C_1(t)-C_5(t)$ are defined in the
supplementary material [\citet{Supp:Stephens3}]. The expansion
suggests that a higher-order accurate quantile of the distribution of
the test statistic may be found by solving for $Z_\alpha^{*}$ such that
$P(T<Z_\alpha^{*})=1-\alpha/2$ for two-sided tests. A~significance test
may therefore be completed by comparing $T_s$ to the corresponding
percentile of the standard normal distribution or to the reference
value determined by \citeauthor{Bickel78}. We refer to the former as
the Approximate Exact Test and the latter as Approximate Exact Test
(BZ).

\subsubsection{Method \textup{IVa}: Exact test}
The exact test also applies to the strong null hypothesis of no
intervention effect on any community's mean response ($H_0\dvtx
y_{a}=y^{*}$ for all $a$); the null distribution of $T_p=S$ is
calculated by permuting the intervention assignment $A_i$ among
subjects. For each permutation, the test statistic $T_p$ is calculated
under the permuted intervention assignment $A_b$, creating the
distribution of statistics $T_p(A_b)$. The exact null distribution is
often estimated by conducting $B$ permutations for large $B$, and a $p$-value
is obtained by $p_B=\frac{1}{B}\sum_{b=1}^{B} I(|T_p(A_b)|>|T_p|)$. For
a level $\alpha$ test, we reject the strong null of no intervention
effect when $p_B<\alpha$. Exact tests are also available in standard
statistical software packages.

\subsection{Dependent outcomes}\label{subsection:Dep}
Averaging individual-level outcomes into a
com\-munity-level statistic may result in loss of information. More
efficient tests of the Young Citizens intervention effect take into
account the possibility of correlation among individual-level responses
of community members. Below, we consider modifications of the
univariate tests that accommodate such correlation.

\subsubsection{\texorpdfstring{Method \textup{Ib}: Wald test of $\beta_1^*$ in the conditional
treatment model using GEE [\citet{Zeger}]}
{Method \textup{Ib}: Wald test of beta1* in the conditional
treatment model using GEE [Zeger and Liang (1986)]}}

To account for correlation in survey responses within a community,
generalized estimating equations may be constructed assuming the CMM.
For individual-level analyses, the weak null hypothesis is that the
average individual response is identical for individuals in communities
assigned to intervention or control, conditional on covariates. The
adjusted intervention effect $\beta_1^*$~is estimated by solving the
generalized estimating equations
%
%
\begin{equation}
\sum_{i=1}^n{\mathbf{D}_i{
\mathbf{V}}_i^{-1} \bigl[\mathbf{Y}_i-\mathbf
{g}(A_i,\mathbf{X}_i;\beta) \bigr]}=\mathbf{0},
\end{equation}
where $\mathbf{D}_i=\frac{d\mathbf{g}(A_i,\mathbf
{X}_i;\beta)}{d\beta^{\mathrm{T}}}$ and
$\mathbf{V}_i=V_i(\phi)^{1/2}\mathbf{R}V_i(\phi)^{1/2}$. The working
covariance $\mathbf{V}_i$~is determined by the $m_i$ by $m_i$
correlation matrix $\mathbf{R}$ and diagonal variance matrix
$V_i(\phi)$. To evaluate $H_0\dvtx E[\mathbf{Y}_i\mid
\mathbf{X}_i,A_i=1]=E[\mathbf{Y}_i\mid \mathbf {X}_i,A_i=0]$, the
standardized coefficient $T_c$ is calculated using the sandwich
variance estimator,
%
%
\begin{eqnarray}\label{eq:sandwichvar}
\qquad\widehat{\operatorname{Var}}(\hat{\beta}) &=& \Biggl(\sum
_{i=1}^n \mathbf{D}_i{
\mathbf{V}}_i^{-1} \mathbf{D}_i
\Biggr)^{-1}
\nonumber\\[-8pt]\\[-8pt]
&&{}\times  \Biggl(\sum_{i=1}^n
\bigl[\mathbf{D}_i{\mathbf {V}}_i^{-1} \bigl\{
\mathbf{Y}_i- \mathbf{g}(A_i,\mathbf{X}_i;
\beta) \bigr\} \bigr]^{\otimes2} \Biggr) \Biggl(\sum
_{i=1}^n \mathbf{D}_i{
\mathbf{V}}_i^{-1} \mathbf{D}_i
\Biggr)^{-1},\nonumber
\end{eqnarray}
which may be calculated in most standard software by requesting a
robust variance and supplying a cluster identifier.

\subsubsection{\texorpdfstring{Method \textup{IIb}: Wald test of $\beta_1$ in the marginal
treatment model using augmented GEE [\citet{Stephens}, \citet{Tsiatis08b}]}
{Method \textup{IIb}: Wald test of beta1 in the marginal
treatment model using augmented GEE
[Stephens, Tchetgen Tchetgen and De Gruttola (2012a),
Zhang, Tsiatis and Davidian (2008)]}}

Assuming the marginal treatment model, augmented estimating equations
are formed by
\begin{eqnarray*}
&& \sum_{i=1}^n{\psi_{a}(O_i;\beta,\eta)}
\\
&&\qquad =\sum_{i=1}^n \Biggl\{
\mathbf{D}_i{\mathbf{V}}_i^{-1} \bigl\{
\mathbf{Y}_i-\mathbf{g}(A_i;\beta) \bigr\}
\\
&&\hspace*{50pt}{} - \sum
_{a=1}^K \bigl\{I(A_i=a)-
\pi_a \bigr\} \bigl[\mathbf{D}_i(a)\mathbf
{V}_i^{-1}(a) \bigl\{\mathbf{d}(\mathbf{X}_i;
\eta_a)-\mathbf{g}(a;\beta) \bigr\} \bigr] \Biggr\}=\mathbf{0},
\end{eqnarray*}
where $\mathbf{d}(\mathbf{X}_i;\eta_a)$ models $E[\mathbf {Y}_{i}\mid
A_i=a,\mathbf{X}_{i}]$. The variance of $\hat{\beta}$ is estimated by
replacing the standard estimating function with the augmented
estimating function $\psi_{a}$ in the middle term of
(\ref{eq:sandwichvar}). Using this method, we test the weak null
hypothesis of equal average responses of individuals randomized to
intervention or control, marginalizing over baseline covariates.

\subsubsection{Method \textup{IIIb}: Approximation to the exact test (multivariate)}

Although child efficacy scores and baseline covariates are considered
fixed for randomization inference, the calculated covariance among
responses in a common community incorporates information on the
difference in the between versus within sum of squares, which may
increase power of tests. A~working covariance $\mathbf{V}_i$ as for
GEE is incorporated into the test statistic given by
%
%
\begin{equation}
\label{eq:Apermmulti} S_D=\sum_{i=1}^n{(A_i-
\pi)\mathbf{1}\mathbf{V}_i^{-1}\mathbf{w}_i},
\end{equation}
where $\mathbf{w}_i$ is the observed value of the residual vector
$\mathbf{W}_i=(W_{i1},W_{i2},\ldots,\break W_{i{m_i}})^{\mathrm{T}}$
determined by $W_{ij}=\hat{\varepsilon}_{ij}=Y_{ij} -
d(\mathbf{X}_{ij};\hat{\eta})$, and $\mathbf{1}$ is the
$m_i$-dimensional vector of~1s. To estimate correlation parameters,
the method of moments is used, as proposed in standard GEE. For
vector-valued outcomes $\mathbf{Y}_i$, the variance of $S_D$ is
%
%
\begin{eqnarray}
\qquad\operatorname{Var}(S_D\mid \mathbf{Y}_i,
\mathbf{X}_i) &=& \pi(1-\pi)\sum_{i=1}^n{
\bigl(\mathbf{1} \mathbf{V}_i^{-1}\mathbf{w}_i
\bigr)^{\otimes2}}
\nonumber\\[-8pt]\\[-12pt]
&&{}+ \biggl(\pi\frac{n/2-1}{n-1}-\pi^2 \biggr)
\overbrace{\sum_{i\neq i'}{ \bigl(\mathbf{1}
\mathbf{V}_{i}^{-1}\mathbf{w}_i \bigr) \bigl(
\mathbf{1}\mathbf{V}_{i'}^{-1}\mathbf{w}_{i'}
\bigr)^{\mathrm{T}}}}^{Q*},\nonumber
\end{eqnarray}
where $Q*$ is the small-sample correction for fixed treatment
allocation. It can be shown that under unequal cluster sizes, if
cluster size is associated with intervention assignment, $E[S_D]\neq
0$, even under the null hypothesis. Type~I error may be preserved under
variable cluster size by mean centering outcomes. The method of
\citeauthor{Bickel78} (\citeyear{Bickel78}) may be applied to dependent
outcomes as well to ensure nominal type~I error levels in small
samples. The null hypothesis tested by this method is that the intervention
has no effect on any individual's response.

\subsubsection{Method \textup{IVb}: Exact test (multivariate)}

The null distribution of test statistic (\ref{eq:Apermmulti}) is
determined by permuting the community-level intervention assignment
$A_i$. Because exact inference conditions on responses and covariates,
the residuals
$\hat{\varepsilon}_{ij}=Y_{ij}-d(\mathbf{x}_{ij};\hat{\eta})$ and
working covariance $\mathbf{V}_i$ do not depend on the permuted
intervention assignment under $H_0$. Working covariance parameters
therefore only need to be estimated once, and $\mathbf{V}_i$ is equal
for all permutations. Testing is conducted as in
Section~\ref{subsection:Ind}.

\subsection{Model selection for baseline covariates}\label{subsection:selection}
When the dimension of the set of baseline covariates is high relative
to sample size, adjusting for all available covariates may be
inefficient. Prior knowledge may suggest the inclusion of some
covariates; for example, the number of children in survey respondents'
households may impact their perception of child efficacy. Among other
covariates whose impact on child efficacy is not well understood, such
as household wealth or ownership of transportation, model selection may
help to determine which covariates to include. Adjusted mean models and
augmented estimation require a conditional mean model that includes
intervention, whereas randomization inference requires that
intervention is left out of the adjustment. A wide array of methods for
selection of baseline covariates is available, particularly for
univariate outcomes. Stepwise selection procedures based on some entry
criterion may be used. Methods based on penalized likelihoods such as
LASSO [\citet{Tibshirani}], adaptive LASSO [\citet{Zou}],
SCAD [\citet{Fan}] and MC$+$ [\citet{Zhang2010}] also apply.
Model selection for multivariate outcomes is less well developed, but
extensions of available methods are presented and discussed in
\citet{Tamar}. We consider two popular approaches, forward
selection by AIC or BIC, and adaptive LASSO, where the tuning parameter
is selected by cross validation, to identify correlates of child
efficacy.

Forward selection is an example of a greedy algorithm, defined as one
that makes the locally optimal choice at each stage in search of a
global optimum [\citet{Black}]. To find the best predictive model,
forward selection starts with a generalized linear model containing the
intercept, and at each step enters a single covariate according to a
prespecified criterion. Examples of entry criteria include minimizing
$p$-values or an information criterion such as AIC, or maximizing
adjusted $r^2$.

Model selection by penalized regression minimizes an objective function
%
%
\begin{equation}
\Omega(\beta)=\sum_{i=1}^nL \bigl
\{Y_i,g(A_i,\mathbf{X}_i;\beta) \bigr\} +
P_{\lambda}(\beta),
\end{equation}
consisting of a loss function $L\{Y_i,g(A_i,\mathbf{X}_i;\beta)\}$ and
a penalty $P_{\lambda}(\beta)$, which is indexed by a nonnegative
tuning parameter $\lambda$. The form of $P_{\lambda}(\beta)$ defines
various regularized regression methods; for adaptive LASSO
$P_{\lambda}(\beta)=\lambda\sum_{k=1}^{p}\hat{w}_k|\beta_k|$ with
weights $\hat{w}_k=1/|\hat{\beta}{}_k^{\gamma}|$ derived from an
initial fit of $\beta$. We consider an adaptive LASSO-hybrid
implementation motivated by the LASSO--OLS hybrid [\citet{Efron}],
in which LASSO is used to determine predictive covariates and the
selected model is subsequently fit by OLS.\looseness=-1

For vector-valued outcomes, \citet{Tamar} suggest that accounting
for correlation improves the efficiency of penalized regression
estimates. In small samples like Young Citizens, it is especially
desirable to reduce the variability in penalized regression, as the
number of units may not be sufficient to achieve consistency despite
estimation under a misspecified independence correlation structure. We
recommend scaling outcomes and covariates by $\Lambda^{1/2}$, where
$\Lambda=\mathbf{V}_i^{-1}$ is a working precision matrix based on an
initial estimate of the coefficient vector. This initial estimate may
be determined by a model selection method that assumes independence.
For validation-based penalized regression, estimation proceeds as in
the univariate case on the scaled outcomes
$\tilde{\mathbf{Y}}_i=\Lambda^{1/2}\mathbf{Y}_i$ and covariates
$\tilde{\mathbf{X}}_i=\Lambda^{1/2}\mathbf{X}_i$. The community and
individual level covariates selected by each method are discussed in
the results.

%
\begin{table}
\tablewidth=245pt
\tabcolsep=0pt
\caption{Average number of baseline covariates selected by AIC, BIC and
Adaptive LASSO by sample size when candidate models include the correct
model. First entry---number of baseline covariates selected when
treatment was forced into the model. Second entry---number of
baseline covariates when treatment was omitted from the model}\label{table:dims}
%
\begin{tabular*}{245pt}{@{\extracolsep{\fill}}ld{2.2}cc@{}}
\hline
$\bolds{n_a}$ & \multicolumn{1}{c}{\textbf{AIC}} & \textbf{BIC} & \textbf{A. LASSO}
\\
\hline
\phantom{0}10 & 10.45 & 7.51 & 6.09 \\
& 8.27 & 5.71 & 4.70
\\[3pt]
\phantom{0}15 & 13.32 & 7.94 & 7.48 \\
& 11.08 & 6.29 & 6.13
\\[3pt]
\phantom{0}25 & 10.71 & 6.74 & 7.10 \\
& 9.38 & 5.44 & 5.73
\\[3pt]
\phantom{0}50 & 10.62 & 6.56 & 7.26 \\
& 9.40 & 5.39 & 5.84
\\[3pt]
100 & 11.10 & 6.80 & 7.41 \\
& 9.96 & 5.71 & 5.92\\
\hline
\end{tabular*}
\end{table}
%

\section{Results: Young Citizens}\label{section:App}

We first present analyses of the Young Citizens study using the
independent outcome methods of Section~\ref{subsection:Ind} and then
repeat the analysis with the dependent outcome methods of
Section~\ref{subsection:Dep}. For the independent outcome analysis,
within-community responses are averaged into a single mean community
score. In the dependent outcome analysis, working independence and
exchangeable working correlations are used to account for correlation
in community members' responses. The presentation closes with a
comparison of the results obtained using each strategy.

\subsection{Independent outcomes}
As stated in Section~\ref{section:Methods}, one strategy for analyzing
clustered data involves averaging individual-level data by cluster and
then employing methods for independent data. In Young Citizens, 30
independent observations were obtained by averaging child efficacy and
baseline covariates by community. Nominal covariates such as ethnic
group and religion were first converted to a set of individual-level
binary variables, each denoting whether or not a subject belonged to a
particular group. The binary indicators were then averaged within
communities to calculate each community's percentage of subjects
falling into each nominal level. We first describe the results of the
AIC, BIC and adaptive LASSO model selection procedures in identifying
baseline covariates that predict child efficacy and then follow with
the primary intervention analysis.

Because of the small number of observations, only main effects were
considered for the covariate-adjusted mean model. For prediction of the
cluster-level \mbox{averages,} forward selection by AIC selected percent Asian
ethnicity (Asian), percent employed (employed), percent knowing the
local leader (leader), average years \mbox{residing} in the community (years),
urban community status (urban), percent self identifying as protestant
(religion\_2) and percent with 6--9 local relatives (relatives\_4). BIC
selected Asian, employed, leader and years. Adaptive LASSO
\mbox{selected} years, percent of surveyed households with a good floor (floor), Asian,
employed and urban. Model selection was repeated for randomization
tests with the omission of treatment from considered models. AIC then
selected years, floor, percent with 3--5 local relatives
(relatives\_3), percent owning transportation (transportation), flush
and percent owning their home (home). The BIC-based model contained
years, floor, relatives\_3 and employed. Adaptive LASSO did not select
any covariates. The predictive power of covariates varied across models
ranging from $r^2={}$0.72--0.82 for models including treatment as a
covariate and $r^2={}$0.48--0.64 for models excluding treatment.

%
\begin{table}
\tabcolsep=0pt
\caption{Analysis of the Young Citizens study: independent,
cluster-averaged. Covariate-adjusted method (Method), regression (R)
\{AIC, BIC, Adaptive LASSO (A. LASSO)\}, test statistic (T) and
$p$-value~($p$), with each test statistic evaluated under independence (Ind) and
exchangeable (Exch) working covariance. $p$-values for Approx. Exact
tests are calculated under Bickel's c.d.f. for randomization test
statistics. ``Unadjusted'' denotes the unadjusted test}\label{table:Analysis2}
\begin{tabular*}{\tablewidth}{@{\extracolsep{\fill}}lccccd{2.4}@{}}
\hline
\textbf{Method} & \textbf{Adjustment} & \textbf{Test statistic} & \textbf{S.t.d. error} & \textbf{$\bolds{Z}$-value} & \multicolumn{1}{c@{}}{$\bolds{p}$}\\
\hline
CMM & AIC & 0.413 & 0.064 & 6.450 & {<}0.0001\\
& BIC & 0.460 & 0.072 & 6.369 & {<}0.0001\\
& LASSO & 0.411 & 0.072 & 5.696 & {<}0.0001\\
& Unadjusted & 0.362 & 0.087 & 4.171 & 0.0003
\\[3pt]
Augmented & AIC & 0.413 & 0.053 & 7.774 & {<}0.0001\\
& BIC & 0.460 & 0.062 & 7.448 & {<}0.0001\\
& LASSO & 0.411 & 0.061 & 6.720 & {<}0.0001
\\[3pt]
Approx. Exact & AIC & 1.358 & 0.487 & 2.787 & 0.0053\\
& BIC & 1.711 & 0.587 & 2.915 & 0.0036\\
& LASSO & 2.713 & 0.814 & 3.334 & 0.0009\\
& Unadjusted & 2.713 & 0.814 & 3.334 & 0.0009
\\[3pt]
Approx. Exact (BZ) & AIC & 1.358 & 0.487 & 2.787 & 0.0045\\
& BIC & 1.711 & 0.587 & 2.915 & 0.0031\\
& LASSO & 2.713 & 0.814 & 3.334 & 0.0005\\
& Unadjusted & 2.713 & 0.814 & 3.334 & 0.0005
\\[3pt]
Exact & AIC & 1.358 & -- & -- & 0.0020\\
& BIC & 1.711 & -- & -- & 0.0030\\
& LASSO & 2.713 & -- & -- & 0.0003\\
& Unadjusted & 2.713 & -- & -- & 0.0003\\
\hline
\end{tabular*}
\end{table}

Results of the cluster-level analysis are shown in
Table~\ref{table:Analysis2}. All methods suggest that the intervention
significantly increases child efficacy. Augmented tests were highly
significant at the $p=0.05$ level, but as shown in the following
section, these methods generally do not preserve type~I error in small
samples. Unadjusted and covariate-adjusted randomization tests also
provided strong evidence of an intervention effect, with smaller
standard errors reported for covariate-adjusted tests than the
unadjusted test.

\subsection{Dependent outcomes}
For CMM and augmented approaches, which include treatment assignment in
the adjustment model, covariates selected by AIC include employed, age,
presence of flushing toilet (flush), number of relatives in the
neighborhood (relatives), religion, transportation, and home at the
individual level and community population density (density) at the
community-level. BICn selected the same covariates as AIC except for
transportation and home. BIC penalized by the number of total
observations (BICm) chose individual-level covariates employed, age and
flush. Finally, adaptive LASSO also chose employed,\vadjust{\goodbreak} age, flush and
religion. For randomization tests, the AIC-based model contained
employment, flush, age, religion, relatives, home and wealth deviance
for each family from the mean community wealth. BICn selected
employment, flush, age, religion and relatives. Selection by BICm and
adaptive LASSO again chose employed, flush and age. All covariates
selected for randomization analyses were individual-level. The
predictive power of covariates ranged from $r^2={}$0.075--0.106 for
models that included treatment and $r^2={}$0.052--0.064 for those that
did not. In unadjusted randomization tests, outcomes were mean centered
as suggested in Section~\ref{subsection:Dep} to preserve type~I error
when cluster size is associated with intervention; in Young Citizens,
intervention communities had on average nine more individuals than
control communities.

%
\begin{table}
\tabcolsep=0pt
\caption{Analysis of the Young Citizens study: dependent.
Covariate-adjusted method (Method), regression (R) \{AIC, BIC by $n$
(BICn), BIC by $M$, (BICm), Adaptive LASSO (A. LASSO)\}, test statistic
(T) and $p$-value ($p$), with each test statistic evaluated under
independence (Ind) and exchangeable (Exch) working covariance. $p$-values
for Approx. Exact tests are calculated under Bickel's c.d.f. for
randomization test statistics. ``Unadjusted'' denotes the unadjusted test} \label{table:Analysis}
%
\begin{tabular*}{\tablewidth}{@{\extracolsep{\fill}}lcd{2.3}d{2.3}cd{2.4}d{2.3}d{2.4}d{1.4}d{2.4}@{}}
\hline
& & \multicolumn{4}{c}{\textbf{Independence}} & \multicolumn{4}{c@{}}{\textbf{Exchangeable}}\\[-6pt]
& & \multicolumn{4}{c}{\hrulefill} & \multicolumn{4}{c@{}}{\hrulefill}
\\
 & \textbf{Adjust-} & \multicolumn{1}{c}{\textbf{Test}}
& \multicolumn{1}{c}{\textbf{S.t.d.}} &
&  & \multicolumn{1}{c}{\textbf{Test}}
& \multicolumn{1}{c}{\textbf{S.t.d.}}&
&
\\
\textbf{Method} & \textbf{ment} & \multicolumn{1}{c}{\textbf{statistic}} & \multicolumn{1}{c}{\textbf{error}}
& \multicolumn{1}{c}{\textbf{$\bolds{Z}$-value}} & \multicolumn{1}{c}{$\bolds{p}$}&
\multicolumn{1}{c}{\textbf{statistic}}& \multicolumn{1}{c}{\textbf{error}} &  \multicolumn{1}{c}{\textbf{$\bolds{Z}$-value}}
&\multicolumn{1}{c@{}}{$\bolds{p}$}
\\
\hline
CMM & AIC & 0.364 & 0.069 & 5.293 &{<}0.0001 & 0.365 & 0.067 & 5.276 &{<}0.0001\\
& BICM & 0.363 & 0.068 & 5.333 &{<}0.0001 & 0.363 & 0.069 & 5.275 &{<}0.0001\\
& BICN & 0.325 & 0.072 & 4.505 &{<}0.0001 & 0.331 & 0.072 & 4.623 &{<}0.0001\\
& LASSO & 0.325 & 0.072 & 4.505 &{<}0.0001 & 0.331 & 0.072 & 4.623&{<}0.0001\\
& Unadjusted & 0.354 & 0.085 & 4.141 &{<}0.0001 & 0.354 & 0.082 & 4.319&{<}0.0001
\\[3pt]
Augmented & AIC & 0.364 & 0.069 & 5.294 &{<}0.0001 & 0.364 & 0.069 &5.312 &{<}0.0001\\
& BICM & 0.363 & 0.068 & 5.353 &{<}0.0001 & 0.365 & 0.069 & 5.281&{<}0.0001\\
& BICN & 0.325 & 0.070 & 4.640 &{<}0.0001 & 0.330 & 0.330 & 4.657&{<}0.0001\\
& LASSO & 0.325 & 0.070 & 4.640 &{<}0.0001 & 0.330 & 0.330 & 4.657&{<}0.0001
\\[3pt]
Approx.  & AIC & 89.833 & 28.676 & 3.133 & 0.0017 & 37.058 &11.123 & 3.332 & 0.0009\\
\quad Exact & BICM & 91.216 & 29.160 & 3.128 & 0.0018 & 36.096 & 10.779 & 3.349 &0.0008\\
& BICN & 88.809 & 28.443 & 3.122 & 0.0018 & 36.588 & 10.994 & 3.328 &0.0009\\
& LASSO & 88.809 & 28.443 & 3.122 & 0.0018 & 36.588 & 10.994 & 3.328 &0.0009\\
& Unadjusted & 95.288 & 32.403 & 2.941 & 0.0033 & 29.103 & 8.8703 &3.2810 & 0.0010
\\[3pt]
Approx.  & AIC & 89.833 & 28.676 & 3.133 & 0.0017 & 37.058& 11.123 & 3.332 & 0.0008\\
\quad Exact & BICM & 91.216 & 29.160 & 3.128 & 0.0017 & 36.096 & 10.779 & 3.349 &0.0007\\
\quad (BZ) & BICN & 88.809 & 28.443 & 3.122 & 0.0018 & 36.588 & 10.994 & 3.328 &0.0009\\
& LASSO & 88.809 & 28.443 & 3.122 & 0.0018 & 36.588 & 10.994 & 3.328 &0.0009\\
& Unadjusted & 95.288 & 32.403 & 2.941 & 0.003 & 29.103 & 8.8703 &3.2810 & 0.0010
\\[3pt]
Exact & AIC & 89.833 & \multicolumn{1}{c}{--} & \multicolumn{1}{c}{--} & 0.0003 & 37.057 & \multicolumn{1}{c}{--} & \multicolumn{1}{c}{--} & 0.0003\\
& BICM & 91.912 & \multicolumn{1}{c}{--} & \multicolumn{1}{c}{--} & 0.0007 & 36.508 & \multicolumn{1}{c}{--} & \multicolumn{1}{c}{--} & 0.0003\\
& BICN & 88.809 & \multicolumn{1}{c}{--} & \multicolumn{1}{c}{--} & 0.0007 & 36.588 & \multicolumn{1}{c}{--} & \multicolumn{1}{c}{--} & 0.0007\\
& LASSO & 88.809 & \multicolumn{1}{c}{--} & \multicolumn{1}{c}{--} & 0.0007 & 26.588 & \multicolumn{1}{c}{--} & \multicolumn{1}{c}{--} & 0.0007\\
& Unadjusted & 95.288 & \multicolumn{1}{c}{--} & \multicolumn{1}{c}{--} & 0.0010& 29.103 & \multicolumn{1}{c}{--} & \multicolumn{1}{c}{--} & 0.0003\\
\hline
\end{tabular*}
\end{table}

Table~\ref{table:Analysis} presents the Young Citizens individual-level
analysis. Adjusted and augmented GEE methods were associated with
highly significant treatment effects $(p<0.0001)$ across
covariate-adjusted and unadjusted tests. For the approximate exact
tests, covariate-adjusted and unadjusted methods yielded a significant
intervention effect with either correlation structure. Applying
Bickel's small-sample adjustment to obtain tail probabilities resulted
in $p$-values that were slightly larger than those based on the
standard normal distribution. Significant intervention effects were
also detected using exact tests with either working covariance
structure. The value of baseline covariate adjustment is shown in
examining the approximate exact test under the independence model,
where standard errors decreased for covariate-adjusted vs. unadjusted
tests. Under the exchangeable correlation structure standard errors
were larger for covariate-adjusted tests than unadjusted tests.
Altogether, the data provide sufficient evidence that children living
in communities that had received the intervention were perceived as
more knowledgeable and equipped to educate their peers about HIV than
children whose communities did not. The results underscore the
importance of using appropriate methodology. The unadjusted tests
based on GEE methods were highly significant, but, as shown in the
following simulation studies of Section~\ref{section:Sim}, the validity
of such methods is not guaranteed when the number of clusters is fairly
small and no small-sample variance adjustment is used.

\subsection{Comparison of cluster and individual-level analyses}

Although both levels of analyses provide evidence of an intervention
effect, key differences were observed in the results of various methods
between individual-level and cluster-level Young Citizens analyses.
The set of covariates selected by model selection was different for
cluster-level vs. individual-level analysis, with higher $r^2$ values
observed in models for the cluster-level analysis. In cluster-level
analyses, the variance of the test statistic decreased with covariate
adjustment. The variances of covariate-adjusted approximate exact
randomization test statistics were approximately half\vadjust{\goodbreak} of those of the
unadjusted statistic variances. The impact of covariate adjustment on
variance in individual-level analyses varied with choice of working
covariance. When assuming an independence working covariance, the
variances of covariate-adjusted tests were at least 19\% smaller than
the variances of unadjusted tests. Under exchangeable correlation,
covariate adjustment increased variances relative to the unadjusted
test by about 50\%.

\section{Simulation studies}\label{section:Sim}

Simulation studies were conducted to investigate the properties of the
four methods described above in small samples. Section~\ref{sec5.1}
considers methods for independent scalar outcomes that are measured for
each randomized unit, as in the community-averaged Young Citizens
analysis. Following the individual-level Young Citizens analysis,
Section~\ref{sec5.2} provides simulation results for vectors of
dependent outcomes for each randomized group, where methods account for
potential correlation among outcomes within a group. The final
subsection discusses implications for Young Citizens.

\subsection{Independent outcomes}\label{sec5.1}
We first consider scalar outcomes $Y_i$. For each simulated data set 25
baseline covariates $X_{i_1},\ldots, X_{i_{25}}$ were generated from
the multivariate lognormal distribution by exponentiating draws from
the multivariate normal distribution with mean $\mu=(0,0,\ldots,0)$ and
covariance $\Sigma$, where $\Sigma$ was defined such that
$\operatorname{corr}(\log(X_{i_k}),\log(X_{i_{k'}}))=0.5$ for
$k$, $k'=1,\ldots,10$,
$\operatorname{corr}(\log(X_{i_k}),\log(X_{i_{k'}}))=0.2$ for
$k=1,\ldots,10$, $k'=11,\ldots,20$,
$\operatorname{corr}(\log(X_{i_k}),\log(X_{i_{k'}})=0$ for
$k=1,\ldots,20$, $k'=2,\ldots,25$, and
$\operatorname{Var}(\operatorname{log}(X_{i_k}))\hspace*{-0.3pt}=1$ for $k=1,\ldots,25$. Skewed
covariates were generated to ensure that results did not rely on
symmetry, as covariates may not be symmetric in actual data. Treatment
$A_i$ was binary and simulated with a fixed, equal number of subjects
assigned to treatment or control. Outcomes were generated from the
model $Y_i=\eta_0+\eta_1A_i+\eta_2X_{i_1}+\eta_3X_{i_2}+\eta
_4X_{i_{10}}+\eta_5{X_{i_{11}}}\eta_6X_{i_{12}} + \varepsilon_i$ with
$\log(\varepsilon_i)\sim N(0,1.1)$, $\eta'=(1,0,1,1,0.2,0.2,0.2)$ under
the null and $\eta'=(1,4,1,1,0.2,0.2,0.2)$ under the alternative.
Sample sizes of $n_a=10, 15,25,50,100$ in each treatment arm were
considered. Under this design, baseline covariates accounted for 73\%
of the variability in $Y_i\mid A_i$---similar to what was observed in
the Young Citizen's study.

All four covariate-adjusted methods were applied to each simulated data
set, and various adaptive procedures were used to select among the 25
baseline covariates. Several variations for each covariate-adjusted
test were considered, with each variation defined by a different
regression model. For adaptive approaches, selection of regression
models was based on three methods: forward selection minimizing AIC,
forward selection minimizing BIC, and the adaptive LASSO--OLS hybrid.
The adaptive LASSO tuning parameter was selected by $l$-fold
cross-validation, where $l=n/10$. For Method~Ia, inference was
performed by OLS on the model including $A_i$ and covariates suggested
by the adaptive model selection procedure. Adaptively selected models
were compared to two fixed models: the data-generating model, which
serves as a benchmark for the largest possible improvement in power,
and an incorrect model, $E[Y_i\mid \mathbf{X}_i,A_i]=\eta_0 +
\eta_1X_{i_1} + \eta_2X_{i_3} + \eta_3X_{_{10}}+\eta_4X_{i_{13}} +
\eta_5X_{i_{21}}$, that included two predictive covariates and 3 noisy
covariates. We chose to include a fixed covariate-adjusted model to
mirror settings where a select few baseline covariates are known a
priori to correlate with the trial outcome. Finally, each method was
also applied to the unadjusted outcomes $Y_i$ to assess whether
incorporating baseline covariates improved power compared to no
adjustment. Treatment was forced into the regression model for
Methods~Ia~and~IIa. In investigation of Methods~IIIa~and~IVa, treatment
was omitted from covariate selection, as the strong null excludes any
estimated effect of treatment even if not significant. In addition to
assessing type~I error and power when the set of candidate models
included the true data-generating model, we also assessed power when
important transformations of baseline covariates were not included. We
modified the data-generating mechanism to include squared terms for
$X_{i_1}$ and $X_{i_{10}}$ and changed the coefficient of $X_{i_1}$ to
$\eta_1=0.50$. As in the previous setting, model fitting algorithms for
determining predictive covariates only considered linear terms.

%
%
\begin{figure}[!t]

\includegraphics{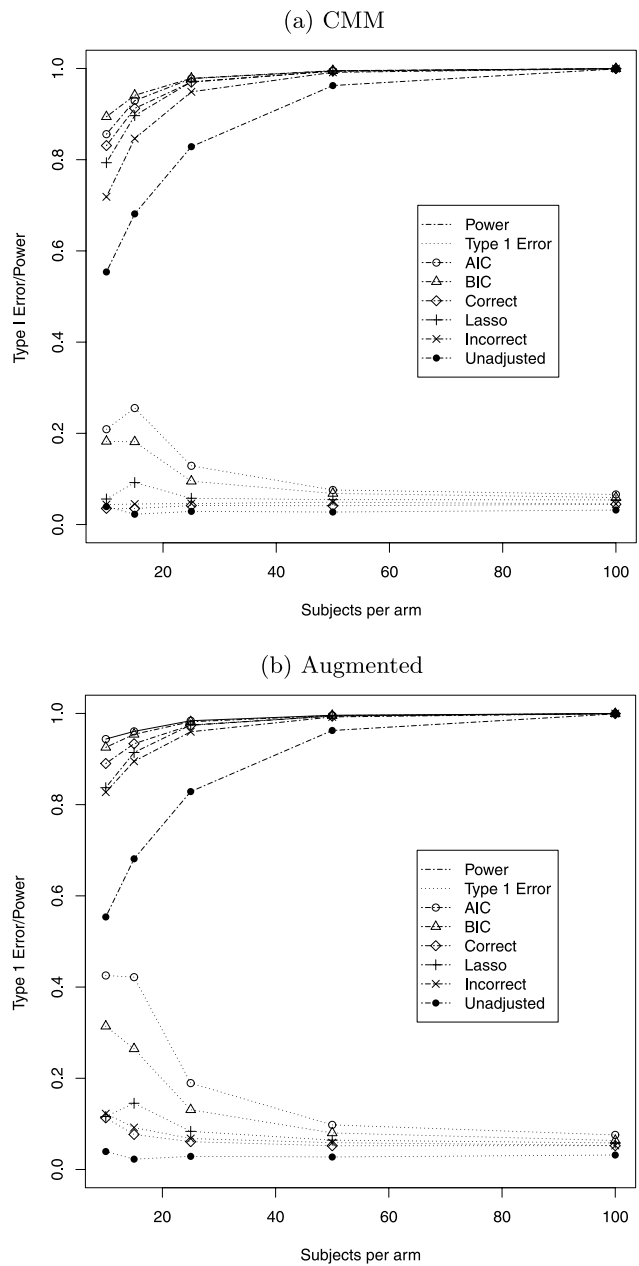}

\caption{Type~I error and power of univariate CMM and augmented tests.
Adaptive regression model selection: AIC, BIC, Adaptive LASSO.
Prespecified models: Correct, Incorrect. ``Unadjusted''
denotes the test statistic that does not incorporate baseline
covariates.}\label{fig:ANCOVAAUG:subfig}
\end{figure}

%
%
\begin{figure}

\includegraphics{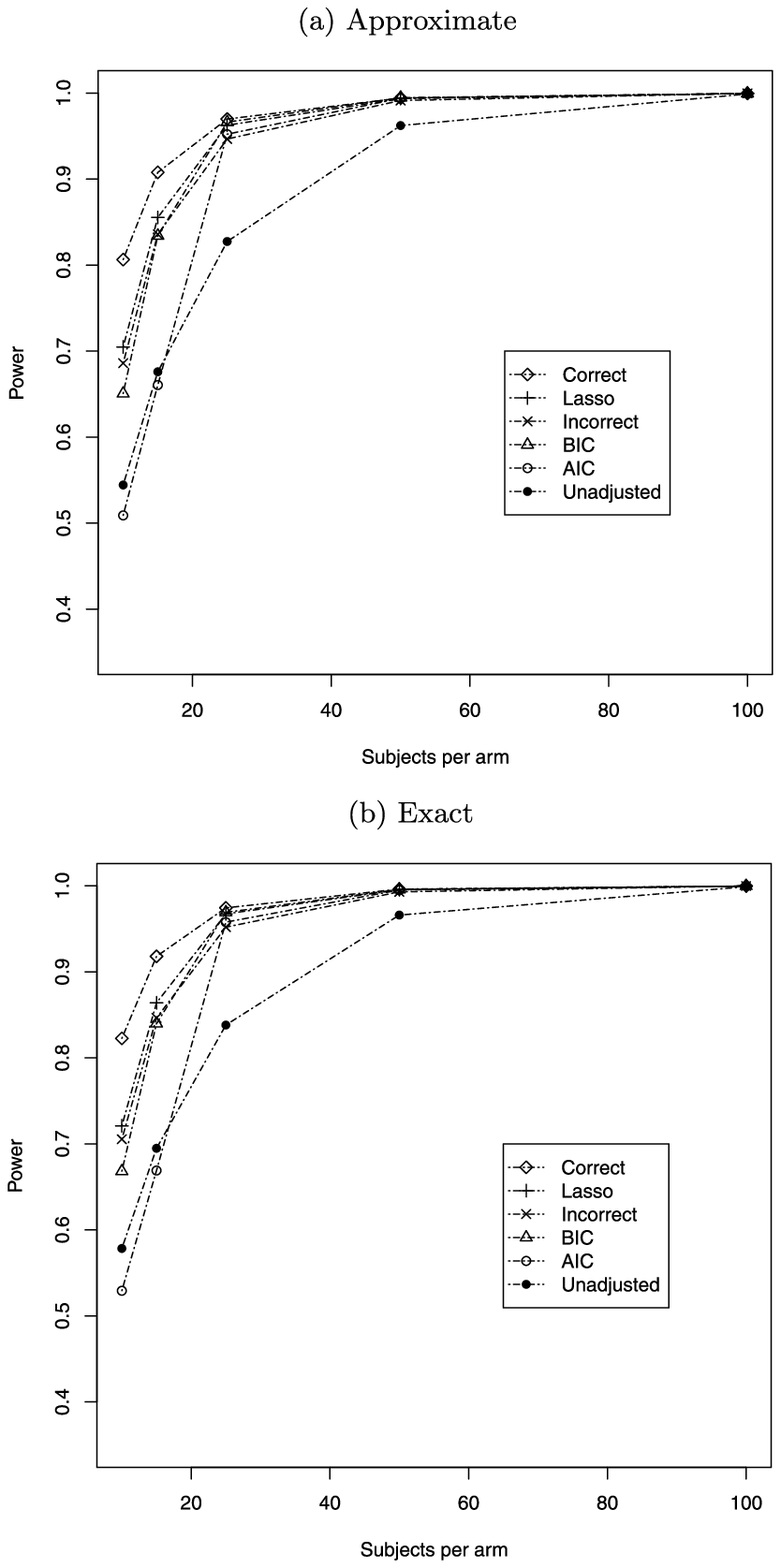}

\caption{Power of univariate approx. exact and exact tests
when the correct model is a candidate model. Adaptive regression model
selection: AIC, BIC, Adaptive LASSO. Prespecified models: Correct,
Incorrect. ``Unadjusted'' denotes the test statistic that does not
incorporate baseline covariates.}\label{fig:Exact:subfig} 
\end{figure}

Results for type~I error are shown below in
Figure~\ref{fig:ANCOVAAUG:subfig} and Table~\ref{table:dims} of the
supplementary material [\citet{Supp:Stephens3}]. All tables
summarizing simulation results are contained in the supplementary
material [\citet{Supp:Stephens3}]. Method~Ia performed poorly for
small sample sizes with model selection, leading to type~I error rates
as large as $\alpha=0.25$. For fixed models chosen a priori, testing
the adjusted treatment effect $\beta^*_1$ preserves type~I error and is
even slightly conservative as a result of the skewness in the
covariates and outcomes ($\alpha={}$0.044--0.049). The performance of
asymptotically equivalent Method~IIa varies over the choice of model
selection procedure. For adaptive LASSO, the augmented test resulted in
type~I errors approximately three times the nominal level at $n_a=10$.
Adaptive selection of covariates by AIC or BIC had even larger type~I
error inflation ($\alpha={}$0.39--0.52 for $n_a=10$). Type~I error was
still not preserved when augmenting with fixed models (0.12 for
$n_a=10$). By contrast, Methods~IIIa~and~IVa maintained type~I error at
all sample sizes considered. The approximate exact test remained
slightly conservative due to the skewness in the data, whereas the
exact test preserved nominal type~I error levels. Regarding model
selection, there were noteworthy differences in the behavior of the
various methods. As expected, BIC favored more parsimonious models than
did AIC; AIC-based selection resulted in models with 9 to 13 baseline
covariates on average; BIC, 6 to 8 covariates. Adaptive LASSO was the
most conservative model selection procedure and included 4 to 7
covariates, with the number of selected covariates increasing with the
sample size. These data are displayed in Table~\ref{table:dims}.

Figure \ref{fig:Exact:subfig} and Table~1 of the
supplementary material [\citet{Supp:Stephens3}] provides
simulation results demonstrating the impact of model selection
procedures on power. For $n_a\leq50$, covariate adjustment based on AIC
and BIC resulted in larger power than did the correct covariate
adjustment model for Methods~Ia~and~IIa (Power${}={}$0.86--0.99 for AIC
and BIC, Power${}={}$0.83--0.99 for the correct model), suggesting that
the former led to overfitting of the regression model. The power of
adjustment with adaptive LASSO did not exceed the power of adjustment
under the correct model for any covariate-adjusted test statistic
considered. In general, Methods~IIIa~and~IVa had lower power than
Methods~Ia~and~IIa, reflecting the fact that the randomization-based
tests preserve type~I error, whereas adding covariates to the mean
model and augmentation tests do not. For very small sample sizes
($n_a\leq15$), covariate adjustment by AIC in randomization tests
resulted in lower power than the unadjusted test (Approx. Exact
AIC${}={}$0.51--0.66, Approx. Exact Unadjusted 0.54--0.68; Exact
AIC${}={}$0.53--0.67, Exact Unadjusted${}={}$0.58--0.70). For $n_a
\geq25$, AIC-based adjustment improved power compared to no adjustment.
Model selection by BIC and adaptive LASSO, which penalize more severely
for model complexity than AIC, improved power over unadjusted test
statistics across all simulated sample sizes. Method~IVa had higher
power than Method~IIIa, with the difference in power increasing
inversely with sample size. Across all settings considered, Bickel's
adjustment for the distribution of the approximate exact test had
little impact on resulting inferences, suggesting that even
higher-order terms may be necessary to preserve nominal type~I error.

\begin{figure}[t!]

\includegraphics{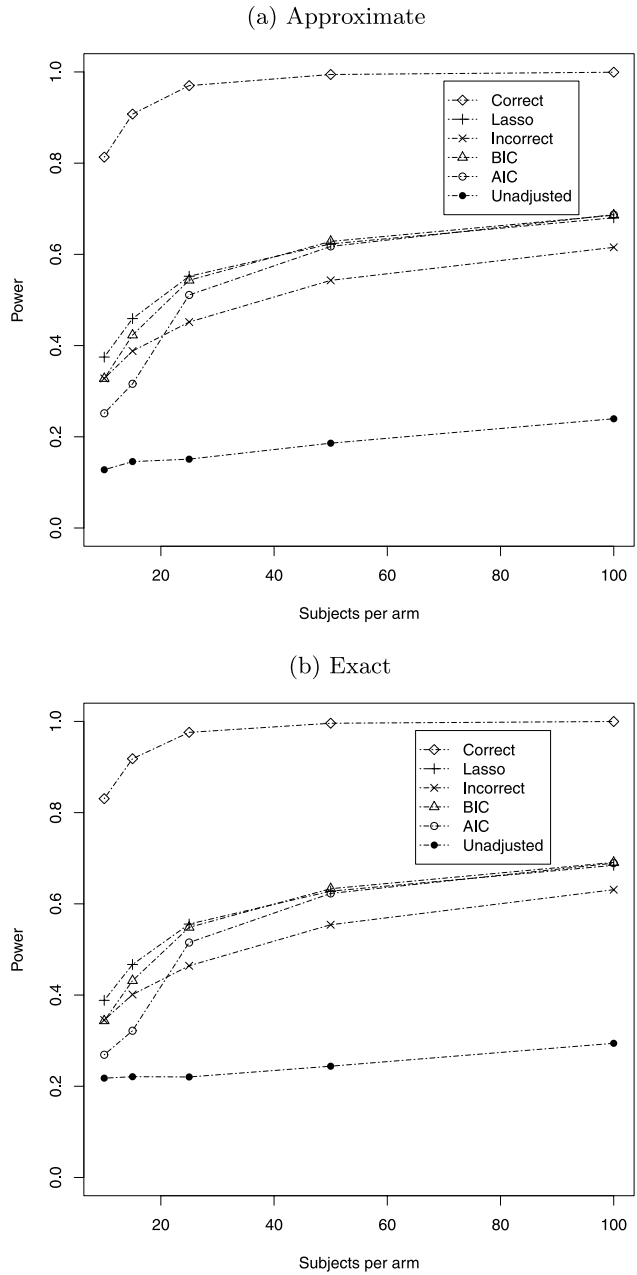}

\caption{Power of univariate approx. exact and exact tests
when the correct model is not a candidate model. Adaptive model
selection: AIC, BIC, Adaptive LASSO. Prespecified models: Correct,
Incorrect. ``Unadjusted'' denotes the test statistic that does not
incorporate baseline covariates.}\label{fig:ExactWro:subfig} 
\end{figure}

In the second set of power simulations, the data-generating model
contained quadratic terms that were not considered in covariate
adjustment. Results are shown in Figure~\ref{fig:ExactWro:subfig} and the
supplementary material [\citet{Supp:Stephens3}]  Table~3. The relative
performance of adaptive procedures remained the same. At small samples
sizes, exact inference AIC resulted in less power improvement than did
the other adjustment methods, but greater power than not adjusting at
all (0.27--0.32 AIC, 0.34--0.47 BIC and adaptive LASSO, 0.34
prespecified incorrect model 0.22 unadjusted). For Method~IIIa, power
gains when AIC was used in the adjustment were again less than those
achieved using BIC selection, adaptive LASSO and the prespecified
incorrect model (AIC${}={}$0.25, Unadjusted${}={}$0.12, BIC${}={}$0.33, adaptive
LASSO${}={}$0.37, Prespecified${}={}$0.33). Increasing the sample size per
arm to $n_a=25$, power for AIC-selected adjustment was more similar to
that associated with BIC and adaptive LASSO. At $n_a\geq50$, all
adaptive procedures resulted in similar power, while the incorrect
prespecified model had lower power (Prespecified${}={}$0.45--0.63,
Adaptive \mbox{Methods${}={}$0.51--0.69}).

%

\subsection{Dependent outcomes}\label{sec5.2}

To evaluate clustered outcome data, values for covariates
$X_{ij_1},\ldots,X_{ij_{25}}$ were generated, with $X_{ij_k}=X_{i_k}$
for $k=1,\ldots,10$. For each cluster,
$(\log(X_{i_1}),\ldots,\log(X_{i_{10}}))\sim MVN(\mathbf{0},\Sigma_2)$,
where $\Sigma_2$ was defined such that
$\operatorname{corr}(\log(X_{i_k}),\log(X_{i_{k'}}))=0.5$ for
$k=1,\ldots,5$, $k'=1,\ldots,5$ and $k=6,\ldots,10$, $k'=6,\ldots,10$,
$\operatorname{corr}(\log(X_{i_k}),\log(X_{i_{k'}}))=0.2$ for $k=1,\ldots,5$,
$k'=6,\ldots,10$. Each covariate $X_{ij_{k}}$ for $k=11,\ldots,20$ was
simulated from the multivariate lognormal distribution with
$\operatorname{corr}(\operatorname{log}(X_{{ij}_k}),\operatorname{log}(X_{{ij'}_k}))=0.2$ independently across $k$.
Finally, for $k=21,\ldots,25$, $\operatorname{log}(X_{ij_{k}})\sim N(0,25)$ with
independence between and within clusters. Binary treatment $A_i$ was
generated with $P(A=1)=0.5$, with the total number of clusters assigned
to each treatment level fixed accordingly. To induce unexplained
correlation within clusters, random cluster effects $b_i$ were
simulated, with $\operatorname{log}(b_i)\sim N(0,\rho\sigma^2)$, where $\rho$ was
varied to induce high or low intracluster correlation. Outcomes
$Y_{ij}$ were generated from the model
$Y_{ij}=\eta_0+\eta_1A_i+\eta_2X_{i_1}+\eta_3X_{{ij}_{11}}+\eta
_4X_{i_3}+\eta_5{X_{{ij}_{12}}}\eta_6X_{{ij}_{15}} +
b_i+\varepsilon_{ij}$, with $\log(\varepsilon_{ij})\sim
N(0,\sigma^2=2.8)$. We set the coefficient vector $\eta
=(1,0,1.25,1.25,0.2,0.2,0.2)$ under the null hypothesis of no treatment
effect, and $\eta=(1,2.2,1.25,1.25,0.2,0.2,0.2)$ under the alternative.
These covariates and $\eta$ values were chosen for the data-generating
model to include at least one strongly predictive and one weakly
predictive covariate at each of the cluster and individual levels.
Monte Carlo data sets consisted of $n=10,15,25$ clusters of size
$m_i=20,30$ or $n=25,50,100$ clusters of size $m_i=4,6,8$ per treatment
arm. The correlation scale parameter was set to $\rho=10/19$, inducing
a conditional correlation [$\operatorname{corr}(Y_{ij},Y_{ij'}\mid
\mathbf{X}_{i},A_i)$] of 5\% and 5.6\% of variability in $Y_{ij}\mid
A_i$ explained by baseline covariates. In Young Citizens, the median
cluster size was $\tilde{m}=31$, intracluster correlation was nearly
5\%, and the $r^2$ of predictive models ranged from 0.052--0.106.
Average cluster size, intracluster correlation and predictiveness of
covariates under the simulation design were therefore similar to Young
Citizens when considering the dependent outcome data structure. In a
second set of simulations we set $\log(\varepsilon_{ij})\sim
N(0,\sigma^2=1.9)$ and $\rho=1$, corresponding to $r^2=0.17$ and a
conditional correlation of 50\% to examine the impact of high
intracluster correlation.

We first adaptively determined predictive models for the mean outcome
conditional on baseline covariates without consideration of correlation
among outcomes within a cluster. We then compared these results to the
Monte Carlo power of adjusted tests when model selection did account
for correlation in responses (Section~\ref{subsection:selection}).
Selection of baseline covariates for adjustment included forward
selection by AIC, two modifications of BIC for multivariate data and
adaptive LASSO. All regression models were ultimately fit by OLS. For
BIC, two regression models were selected: the first considered the
number of clusters in the penalty for model complexity (BICn), and the
second calculated BIC based on the total number of individual-level
observations (BICm).

In deriving BIC for linear mixed models, \citet{Pauler} showed
that for a random intercept model the true penalty is of the form
$\Omega_h=\sum_{k=1}^{p}\operatorname{log}(N_k^*)$, where $h$ indexes
candidate models, $k$ indexes the $p$ covariates in the $h$th model,
$N_k^*=n$ for between-cluster effects, and $N_k^*=M$ for within-cluster
effects. BICm and BICn would therefore correspond to the true BIC for
models containing only cluster-level covariates or individual-level
covariates, respectively. Evaluating the true BIC for models including
both types of covariates requires calculating $\Omega_h$ for each
candidate model in the stepwise procedure by observing its number of
cluster-level and individual-level covariates. To ease computational
burden, BICm and BICn were used. The adaptive LASSO tuning parameter
was selected based on fivefold cross-validation. The two fixed
regression models included the data-generating model and an incorrect
model, $E[Y_{ij}\mid \mathbf{X}_{ij},A_i]=\eta_0 + \eta_1X_{i_1} +
\eta_2X_{i_2} + \eta_3X_{i_{10}}+\eta_4X_{{ij}_{13}} +
\eta_5X_{{ij}_{21}}$, including two predictive covariates and 3 noisy
covariates. For Methods~Ib~and~IIb, treatment was forced into the
regression model; model selection and prespecified models for the
randomization tests omitted treatment. The null distribution of the
observed test statistic under the exact test was determined by
permuting the treatment assignment across clusters $b=1000$ times.
Unadjusted tests were also performed for each method and compared to
covariate-adjusted tests. The impact of incorporating the covariance
structure on randomization tests was evaluated by conducting each test
under both independence and exchangeable correlation structures for
each adjustment model. Specification of a covariance structure for
standard GEE and augmented GEE methods have been evaluated elsewhere
[\citet{WangCarey,Stephens}].

All tables for multivariate simulation results may be found in
supplementary material Supplement~C [\citet{Supp:Stephens3}]. Type~I error for each method is presented in
Tables~4--6.
In small samples ($n_a\leq25$) GEE methods fail to control type~I error
for all covariate-adjusted analyses. Inflation of type~I error reflects
small sample bias in the sandwich variance estimator as well as
additional variance induced by model selection. Under model selection,
type~I error was as large as $\alpha=0.21$ for Method~Ib and
$\alpha=0.253$ for Methods~IIb when there were 10 randomized units per
arm. For 15 or 25 clusters per intervention arm, type~I error inflation
was present but less severe ($\alpha={}$0.057--0.132 for $15 \leq n_a
\leq 25$). When the number of clusters was large ($n_a \geq50$),
nominal type~I error levels of $\alpha=0.05$ were achieved even under
adaptive covariate adjustment. For testing treatment effects, model
selection by AIC resulted in the largest type~I error, followed by the
BIC methods; the adaptive LASSO had the least type~I error inflation.
For the randomization tests, the approximate exact test was generally
conservative across all outcomes. The Bickel adjustment for defining
the rejection region increased type~I error levels of the approximate
exact test closer to the nominal level. The exact test had nominal
type~I error across adaptively-selected and prespecified
covariate-adjusted models.

%
%
\begin{figure}

\includegraphics{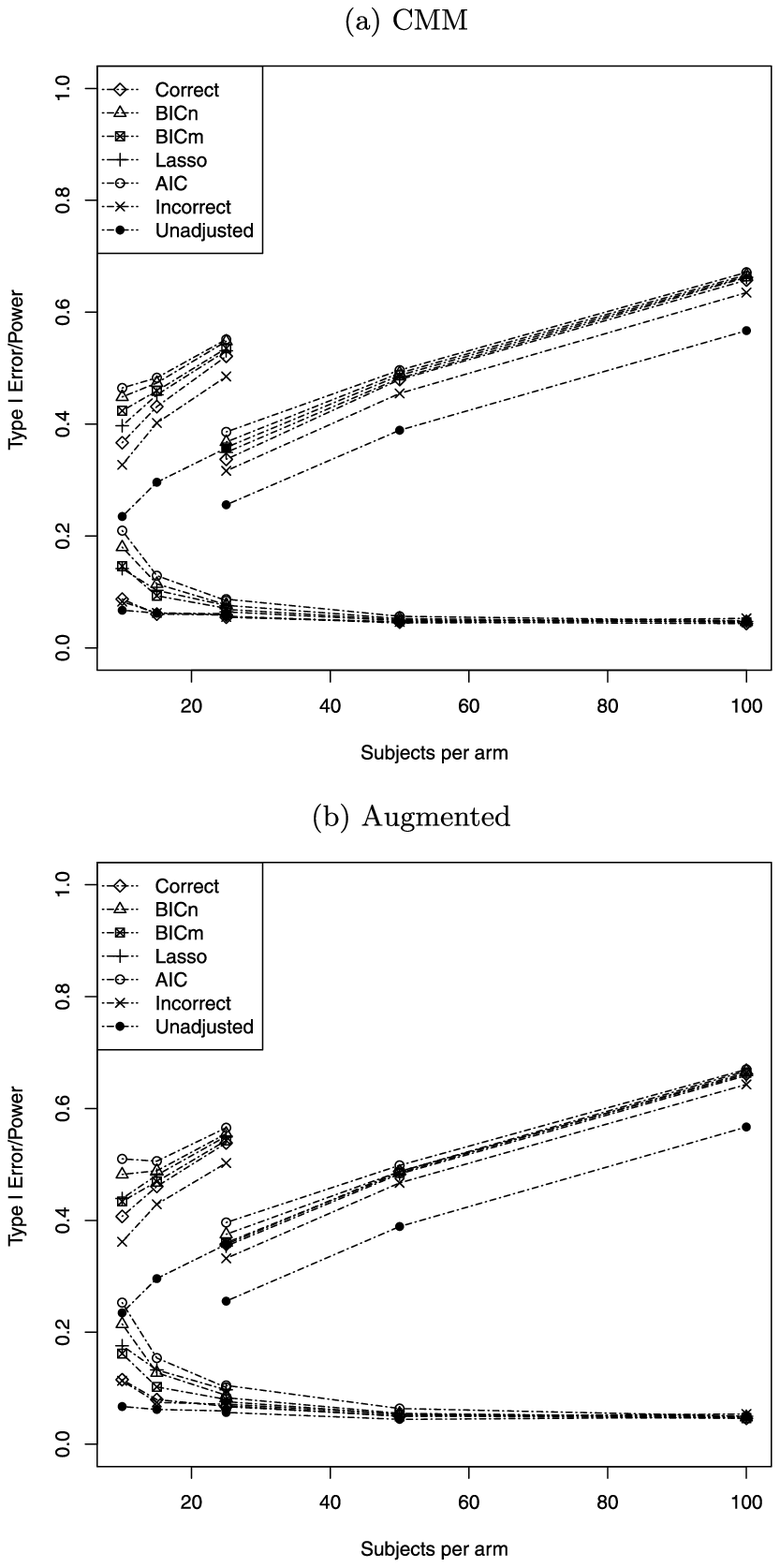}

\caption{Type I error and power of multivariate CMM and
augmented tests. Adaptive regression model selection: AIC, BIC by $n$
(BICn), BIC by $M$, (BICm), Adaptive LASSO (Lasso). Prespecified models:
Correct, Incorrect. ``Unadjusted'' denotes the test statistic that does
not incorporate baseline covariates.}\label{fig:MultAdjAug} 
\end{figure}
%
%
%
\begin{figure}

\includegraphics{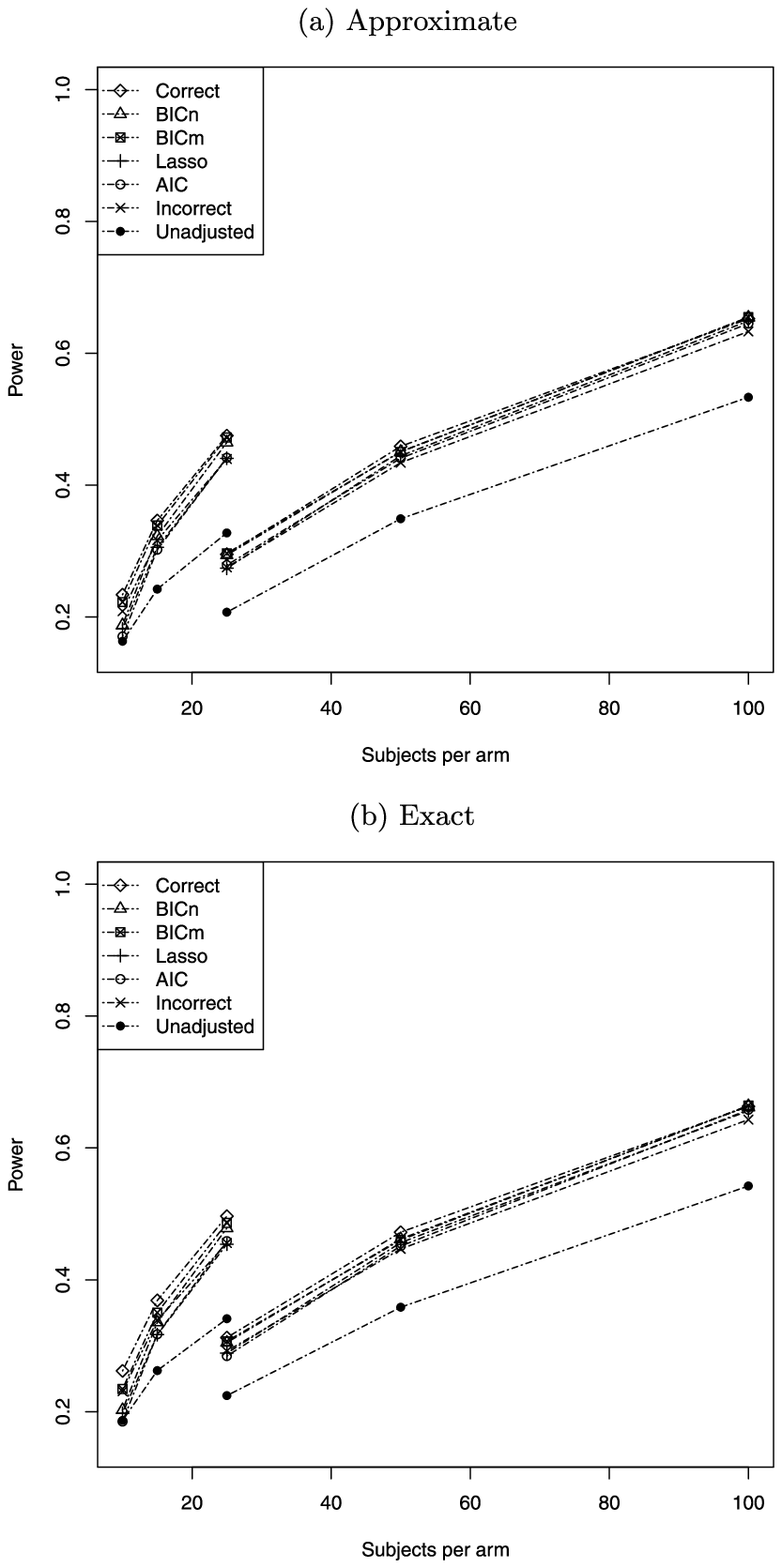}

\caption{Power of multivariate approx. exact and exact tests:
low correlation. Adaptive regression model selection: AIC, BIC by $n$
(BICn), BIC by $M$, (BICm), Adaptive LASSO (Lasso). Prespecified models:
Correct, Incorrect. ``Unadjusted'' denotes the test statistic that does
not incorporate baseline covariates.}\label{fig:MultRandLow} 
\end{figure}
%
%
\begin{figure}[t!]

\includegraphics{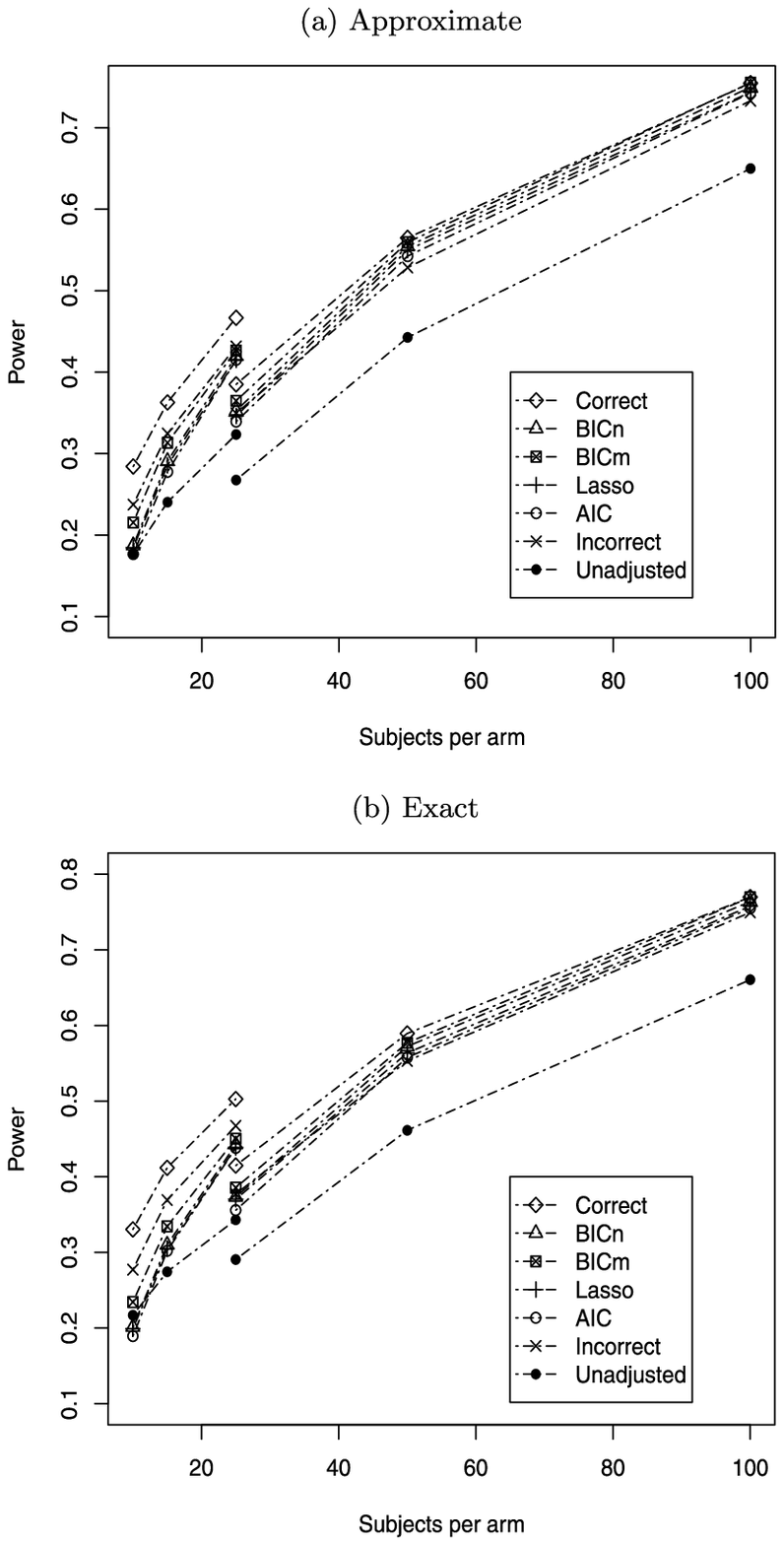}

\caption{Power of multivariate approx. exact
 and exact tests:
high correlation. Adaptive regression model selection: AIC, BIC by $n$
(BICn), BIC by $M$, (BICm), Adaptive LASSO (Lasso). Prespecified models:
Correct, Incorrect. ``Unadjusted'' denotes the test statistic that does
not incorporate baseline covariates.}\label{fig:MultRandHi}
\end{figure}

Figures \ref{fig:MultAdjAug}--\ref{fig:MultRandHi} and Tables~7--12 of the
supplementary material [\citet{Supp:Stephens3}] compare power across covariate-adjusted
tests for dependent outcomes. In most cases, covariate adjustment
improved power compared to the corresponding unadjusted approaches,
regardless of the method of model selection used. Of the adaptive
methods considered, forward selection by BICm resulted in the largest
power for both levels of intracluster correlation. Exchangeable working
covariance specification improved power over working independence only
for randomization tests of the unadjusted outcomes. Method~IVb at $n_a
= 10$ using AIC selection did not improve power over unadjusted
analysis when the exchangeable working covariance was used (Unadjusted
0.187, AIC 0.185). All other model selection techniques resulted in
greater power than unadjusted analyses at all sample sizes and working
covariance structures considered.


\subsection{Implications for Young Citizens}
The simulation study provides insight into inferences for Young
Citizens. In Young Citizens, significant treatment effects were
observed using all four methods of analysis, in both independent and
dependent analyses. Particularly under independent analysis, simulation
studies provide reason to caution against interpreting significant
findings of nonrandomization tests for post-selection inference as
evidence of a treatment effect, considering the severe inflation of
type~I error. In dependent analysis, the type~I error inflation of
nonrandomization tests was also present but not as severe. Significant
findings of the post-selection randomization tests, however, may be
interpreted as evidence of a treatment effect, as randomization tests
preserved type~I error rates after flexible covariate selection. The
decreased standard errors in the Young Citizens analysis for test
statisics that incorporate baseline covariate data compared to
unadjusted test statistics are consistent with simulation results
showing improved power when adjusting for baseline covariates through
both prespecified and adaptive mechanisms.

%
%
\begin{table}
\tabcolsep=0pt
\caption{Characteristics and behavior of covariate-adjusted tests}
\begin{tabular*}{\tablewidth}{@{\extracolsep{\fill}}p{78pt}p{37pt}p{95pt}p{88pt}@{}}
\hline
\textbf{Method} & \multicolumn{1}{c}{$\bolds{H_0}$} & \multicolumn{1}{c}{\textbf{Description}} & \multicolumn{1}{c@{}}{\textbf{Type I error}}\\
\hline
Adjusted mean model\hfill\break \phantom{\quad}(Method Ia--Ib)& \centering{Weak (Neyman)} & Adds baseline covariates to a mean model that contains a treatment variable & Does not preserve type I error under model selection in small samples
\\[3pt]
Augmented\hfill\break \phantom{\quad}(Method IIa--IIb) & \centering{Weak (Neyman)} & Incorporates baseline covariates through a separate augmentation term & Does not preserve type I error under model selection in small samples
\\[3pt]
Approx. Exact\hfill\break \phantom{\quad}(Method IIIa--IIIb) & \centering{Strong (Fisher)} & Tests residuals from a covariate model that does not include treatment and uses the randomization-based variance estimator\vspace*{3pt} & Preserves type I error under model selection in small samples
\\[3pt]
Exact\hfill\break \phantom{\quad}(Method IVa--IVb) & \centering{Strong (Fisher)} & Tests residuals from a covariate model that does not include treatment by permuting the treatment assignment & Preserves type I error under model selection in small samples\\
\hline
\end{tabular*}
\end{table}

\section{Discussion}\label{section:discussion}
We investigated the merits and potential downsides of several
procedures that allow for flexible covariate adjustment when applied to
small samples such as the Young Citizens cluster-randomized trial. We
cannot provide \mbox{guidance} regarding which method optimizes efficiency,
but do provide below a discussion of the precise nature of the null
hypotheses being tested. These hypotheses place restrictions on the
distributions of the outcome, treatment and covariates that may be
ranked from weakest to strongest. The least restrictive (weakest) is
that of the augmented approach, for which the null hypothesis is that
average child efficacy is the same for the two groups of patients
defined by treatment assignment. The null hypothesis tested by the CMM
approach is the next weakest and implies equivalent average child
efficacy among population subgroups---defined by treatment and
additional covariates---regardless of assigned treatment. The
randomization tests consider the same null hypothesis, which is
stronger than that corresponding to CMM and augmented tests. They test
that there is no individual for whom treatment has had an effect; the
null hypothesis being tested is referred to as the strong or sharp null
in distinction to that of the CMM and augmented tests, referred to as
the weak or mean null. Differences at the individual level do not
always imply differences averaged over population subgroups, but
differences at the averaged population level imply differences at the
individual level.

It may be more useful for developing treatment policy to draw
conclusions about average outcomes in subgroups of the population
rather than about variations in individual responses due to treatment.
There may be little interest in promulgating an intervention that
affects individuals but does not reduce the population burden of an
illness. Our investigation demonstrates, however, that there are common
settings for which conclusions drawn about population averages under
flexible covariate selection may be invalid. These settings may be
characterized as having a large number of baseline covariates
considered adaptively for potential power gain and a relatively small
number of randomized units. For univariate outcomes, the augmented
approach, unlike the CMM, resulted in inflated type~I error even under
a prespecified model, reflecting the variability associated with the
nuisance parameters of the conditional model. When responses are
correlated, CMM and augmented methods both suffer from variance
underestimation and type~I error inflation of the sandwich variance
estimator---an occurrence that has previously been noted. For studies
that randomize clusters, the intracluster correlation also affects the
validity of augmented approaches [\citet{Stephens}]. By contrast,
the randomization methods, which condition on outcomes and baseline
covariates, provide valid tests for treatment effects while flexibly
incorporating baseline covariates. The randomization test requires no
assumptions about the underlying data-generating distribution of
outcomes and baseline covariates. As a result, the variability in model
selection of correlates of the outcome does not confer additional
uncertainty in the primary test, thereby preserving type~I error.

Randomization tests provide the most reliable inference when covariate
selection is flexible and sample sizes are small. To provide further
insight into this issue, we consider the interpretation of results from
nonrandomization and randomization tests applied to the same data.
The combination of rejecting $H_0$ using nonrandomization tests and
failing to reject $H_0$ using randomization tests provides evidence
against the validity of the former; if the weak null is properly
rejected, the sharp null cannot hold. By contrast, rejection of the
strong but not the weak null would imply the absence of an average
effect of treatment. Such a scenario would provide support against
rejecting the weak null hypothesis, as the nonrandomization tests are
not generally conservative. Rejection of both tests would provide
evidence for an effect at the individual level; an effect on the
population average is less certain, as a liberal test does not permit
us to distinguish a true positive result from a false positive. Valid
conclusions about population averages would require an unadjusted test
or one in which a select set of covariates are prespecified for
adjustment.

For cluster-randomized studies, the observed trends of type~I error and
power are not expected to vary with individual-level versus
cluster-level covariates. With either type of covariate, flexible
covariate selection will tend to lead to inflated type~I error with
nonrandomization tests, but not with randomization tests. Consideration
of covariates at both levels has important implications for validity
and power of tests. Because the number of randomized clusters may be
small, imbalance among cluster-level characteristics may arise and
distort the interpretation of tests of treatment effects. The same may
hold for individual-level covariates whose distributions vary by
cluster. In the Young Citizens study, there were many more urban than
rural communities, and randomization resulted in an uneven distribution
of the latter, with 6 rural communities in the control arm and 3 in the
intervention arm. To some degree, the effect of urban or rural status
may be mediated through an individual-level covariate such as wealth.
Adjusting for individual level covariates can therefore potentially
reduce the impact of chance imbalance in community characteristics on
test statistics. This is especially relevant when unmeasured community
characteristics impact measured individual-level covariates and
outcomes. Because individual and cluster-level covariates may each
explain variability in the outcome, adjustment will tend to improve
power in testing. As the number of individuals is often much larger
than the number of randomized units, individual-level data provide more
information for a predictive mean model. This is especially true in
settings with small intracluster correlation, which results in a large
effective sample size for individuals.

The dependent data analysis also highlights the interplay between
working covariance selection and baseline covariates. In generalized
linear mixed models for clustered data, random effects are
conceptualized as unmeasured cluster-level covariates that induce
correlation within clusters. A similar rationale applies to
\mbox{unadjusted}
analyses, where imposition of a working covariance structure may
partially account for the impact of baseline covariates on
between-cluster differences, even though these covariates do not
directly enter in the analysis. Unadjusted dependent data methods that
assume working independence ignore the effect of cluster-level and
individual-level covariates whose distributions may vary by cluster on
within-community correlation. Although it is standard practice to
assume working independence and appeal to the robustness of
semiparametric analyses for establishing validity in correlated data
analysis, this strategy may result in inefficient inferences. Our
results suggest that individual-level analyses of clustered data for
the evaluation of intervention effects should include covariate
adjustment or a nonindependence working covariance structure to reduce
residual between-community differences that may mask intervention
effects. If there are \mbox{substantial} between-community differences in
responses, as determined by either unmeasured or \mbox{measured} covariates,
the unadjusted independent data strategy averaging individual-level
data by clusters may be more efficient than unadjusted dependent-data
analyses.

Our investigation also showed that model selection techniques have
varying implications for type~I error and power, depending on the
strength of the penalty used in selecting covariates. The severity of
type~I error inflation varied inversely with penalty strength. Our
discussion of power focuses on randomization tests, as consideration of
power must follow demonstration of validity. Adjustment generally
increased the power of testing for treatment effects over unadjusted
methods, with the caveat that in extremely small samples of independent
outcomes, such as $n_a=10,15$, model selection approaches must be
sufficiently conservative. Our simulation study design used covariates
accounting for 70\% of the outcome variability in independent data and
10\% in dependent data, where the addition of a random cluster effect
diluted the predictive power of covariates. The degree of correlation
between covariates and the outcome impacts the interpretation of our
findings; larger correlation implies greater improvement in power
compared to unadjusted analyses. Model selection by BIC and adaptive
LASSO, which have stronger penalties and therefore favor more
parsimonious models than does AIC, resulted in improved power at the
smallest sample sizes considered. Further research is needed to
formally characterize the power of covariate-adjusted tests under
misspecified covariate adjustment and adaptive covariate
selection.

Our work has focused on hypothesis testing for evaluating treatment
effects; such tests may be inverted to estimate confidence intervals.
When inverting randomization-based hypothesis tests, model selection
needs to be repeated for each potential value of the treatment effect
considered, as estimation of conditional mean models pools across
treated and untreated subjects. Interval estimation may be simplified
by a slight modification of the testing procedure. Under the strong
null, the conditional mean model may be estimated using data only for
untreated subjects. The model may then be applied to all subjects in
conducting the test. Avoidance of pooling the data when estimating the
conditional mean model removes the need for its re-estimation with each
treatment effect value considered. For small-sample univariate data,
it may not be feasible to perform model selection on a single treatment
group, but for a small number of moderately sized clusters such a
strategy may be practicable.

We close with a reference table summarizing the properties of the
flexible covariate-adjusted tests considered.

\begin{supplement}
\stitle{Supplement to ``Flexible covariate-adjusted exact tests of randomized
treatment effects with application to a trial of HIV education''\\}
\slink[doi,text={10.1214/13-AOAS679SUPP}]{10.1214/13-AOAS679SUPP}
\sdatatype{.pdf} \sfilename{aoas679\_supp.pdf}
\sdescription{\textit{Supplement A}: Small sample adjustment of \citeauthor{Bickel78} (\citeyear {Bickel78}).
Function definitions in \citet{Bickel78} small-sample approximation.
\textit{Supplement B}: Simulation study tables---independent outcomes.
Type I error and power of covariate-adjusted tests in independent outcomes.
\textit{Supplement C}: Simulation study tables---dependent outcomes.
Type I Error under low correlation and power under low
correlation and high correlation of covariate-adjusted tests for dependent outcomes.}
\end{supplement}



%

\printaddresses

\end{document}